\documentclass[a4paper,twocolumn,11pt,accepted=2020-04-12]{quantumarticle}
\pdfoutput=1
\usepackage[T1]{fontenc}
\usepackage[latin9]{inputenc}
\usepackage{color}
\usepackage{bm}
\usepackage{xcolor}
\usepackage{amsmath}
\usepackage{amssymb}
\usepackage{graphicx}
\usepackage[numbers,sort&compress]{natbib}
\usepackage{braket}
\usepackage{psfrag}
\usepackage{tikz}
\usepackage[normalem]{ulem}
\usepackage{qcircuit}
\usepackage{subcaption}
\allowdisplaybreaks

\newtheorem{theorem}{Theorem}[section]

\newtheorem{proposition}[theorem]{Proposition}

  {{\nopagebreak\hspace*{\fill}$\Box$\par\vspace{12pt}}}

\newcommand{\cvar}{\ensuremath{\text{CVaR}}}

\begin{document}


\title{Improving Variational Quantum Optimization using CVaR}

\author{Panagiotis Kl.~Barkoutsos}
\affiliation{%
IBM Research -- Zurich
}%
\orcid{0000-0001-9428-913X}

\author{Giacomo Nannicini}
\affiliation{%
IBM T.J.~Watson Research Center
}%
\orcid{0000-0002-4936-1259}

\author{Anton Robert}
\affiliation{%
IBM Research -- Zurich
}%
\affiliation{\'Ecole Normale Sup\'erieure, PSL University, Paris}

\author{Ivano Tavernelli}
\affiliation{%
IBM Research -- Zurich
}%

\author{Stefan Woerner}
\email{wor@zurich.ibm.com}
\affiliation{%
IBM Research -- Zurich
}%
\orcid{0000-0002-5945-4707}



\begin{abstract}
Hybrid quantum/classical variational algorithms can be implemented on noisy intermediate-scale quantum computers and can be used to find solutions for combinatorial optimization problems.
Approaches discussed in the literature minimize the expectation of the problem Hamiltonian for a parameterized trial quantum state.
The expectation is estimated as the sample mean of a set of measurement outcomes, while the parameters of the trial state are optimized classically.
This procedure is fully justified for quantum mechanical observables such as molecular energies.
In the case of classical optimization problems, which yield diagonal Hamiltonians, we argue that aggregating the samples in a different way than the expected value is more natural.
In this paper we propose the Conditional Value-at-Risk as an aggregation function.
We empirically show -- using classical simulation as well as quantum hardware -- that this leads to faster convergence to better solutions for all combinatorial optimization problems tested in our study.
We also provide analytical results to explain the observed difference in performance between different variational algorithms.
\end{abstract}

\maketitle

\section{\label{sec:introduction} Introduction}
Combinatorial optimization (CO) has been extensively studied, because
it is applied in many areas of business and science, and it is
a source of interesting mathematical problems
\cite{nemhauser1988}. Even if many CO problems are NP-hard, they are
routinely solved (possibly not to provable optimality) in industrial
applications. It is widely believed that quantum computers cannot
solve such problems in polynomial time (i.e., NP $\not\subseteq$ BQP
\cite{bernstein1997quantum}), but there is significant effort toward
designing quantum algorithms that could be practically useful by
quickly finding near-optimal solutions to these hard problems. Among
these algorithms, two candidates are more likely to be efficiently
implementable on noisy quantum computers: the
Variational Quantum Eigensolver (VQE)
\cite{moll2018quantum,fried2017qtorch}, and the Quantum Approximate
Optimization Algorithm (QAOA)
\cite{Farhi2014,farhi2017quantum,crooks2018performance}. Both VQE and
QAOA use a parametrized quantum circuit $U(\theta)$ (also called
variational form) to generate trial wave functions $\ket{\psi(\theta)}
= U(\theta)\ket{0}$, guided by a classical optimization algorithm that
aims to solve:
\begin{eqnarray}
\min_{\theta} \bra{\psi(\theta)}H\ket{\psi(\theta)}. \label{eq:vqe_expected_value}
\end{eqnarray}
This expression encodes the total energy of a system through its
Hamiltonian $H$, and more specifically its expected value
$\bra{\psi(\theta)}H\ket{\psi(\theta)}$. There is a well-known
transformation to map CO problems into a Hamiltonian, see e.g.,
\cite{lucas2014ising}; this will also be discussed subsequently in
this paper. Empirical evaluations indicate that VQE's performance
suffers from some key weaknesses in the context of CO problems, and
there is significant room for improvement \cite{Nannicini2019}. At the
same time, the literature on QAOA paints a mixed picture: while some
work concludes that QAOA has promise
\cite{crooks2018performance,niu2019optimizing}, other papers indicate
that it may not perform better than classical algorithms
\cite{hastings2019classical}.

In this paper, we propose a modification to the problem
(\ref{eq:vqe_expected_value}) that is given to the classical
optimization algorithm: rather than minimizing the expected value
$\bra{\psi(\theta)}H\ket{\psi(\theta)}$, we minimize its Conditional
Value-at-Risk (CVaR) -- a measure that takes into account only the
tail of the probability distribution and is widely used in finance
\cite{acerbi2002coherence}. We argue that CVaR more closely matches
the practical goal of heuristic solution methods for CO problems. We
provide an analytical and numerical study, showing that the proposed
methodology significantly improves performance and robustness of VQE
and QAOA in the context of CO.  As a byproduct of this analysis, we
find that for small-depth circuits it is important to use variational
forms with a sufficient number of parameters, e.g., at least linear in
the number of qubits. QAOA for fixed (small) depth does not satisfy
this requirement, and produces ``flat'' quantum states that yield low
probability of sampling the optimal solution.  Thus, not only does our
paper introduce a simple way to improve the performance of the two
most prominent candidates for CO on noisy quantum computers, but it
also sheds some light on their behavior with small-depth circuits.
While we do not claim that this brings their performance on par with
state-of-the-art classical heuristics, our results indicate a marked
improvement as compared to what is proposed in the literature. The
numerical experiments discussed in this paper are implemented in the
open-source library Qiskit \cite{Qiskit}, and executed on classical
quantum simulators as well as IBM's quantum hardware.

The rest of this paper is divided as follows. The next two sections
set the context for this paper by giving an overview of key concepts
from the literature: the mapping of CO problems as a Hamiltonian, VQE
(Section \ref{sec:vqe}), and QAOA (Section \ref{sec:qaoa}). Section
\ref{sec:cvar} formally introduces our main contribution, CVaR
optimization. An analysis of the method is given in Section
\ref{sec:cvar_analysis}. Section \ref{sec:experimental_setup} provides
an empirical evaluation of the proposed method on classically
simulated quantum circuits, and using quantum hardware. Section
\ref{sec:qaoa_analysis} provides a formal analysis of certain aspects
of QAOA to explain the experimentally observed behavior. Finally,
Section \ref{sec:conclusions} concludes the paper.

\section{\label{sec:vqe} Variational Quantum Eigensolver}

The Variational Quantum Eigensolver (VQE) is a hybrid
quantum/classical algorithm originally proposed to approximate the
ground state of a quantum chemical system, i.e., the state attaining
the minimum energy \citep{Peruzzo2014}. This is achieved by solving a
problem of the form (\ref{eq:vqe_expected_value}), with $H$ encoding
the total energy of the quantum chemical system. The same approach can
be used to attempt to solve CO problems. Consider a quadratic
unconstrained binary optimization (QUBO) problem on $n$ variables:
\begin{eqnarray}
\min_{x \in \{0, 1\}^n} b^T x + x^T A x, \label{eq:qubo}
\end{eqnarray}
for given $b \in \mathbb{R}^n$ and $A \in \mathbb{R}^{n \times n}$.
Using the variable transformation $x_i = (1 - z_i)/2$ for $z_i \in
\{-1, +1\}$, problem (\ref{eq:qubo}) can be expressed as an Ising spin
glass model:
\begin{eqnarray}
\min_{z \in \{-1, +1\}^n} c^T z + z^T Q z, \label{eq:ising}
\end{eqnarray}
where $c$ and $Q$ are easily computed from (\ref{eq:qubo}). These two
equivalent problems encompass binary optimization problems, i.e., CO
problems, and they are NP-hard \cite{barahona1982computational}.

Problem (\ref{eq:ising}) can be translated into a Hamiltonian for an
$n$-qubit system: we replace $z_i$ by the Pauli Z-matrix
\begin{eqnarray}
\sigma_Z^i &=&
\left(
\begin{array}{cc}
1 & 0 \\ 0 & -1
\end{array}
\right) \label{eq:sigma_z}
\end{eqnarray}
acting on the $i$-th qubit, and each term of the form $z_i z_j$ by
$\sigma_Z^i \otimes \sigma_Z^j$. The summation of (tensor products of) Pauli terms obtained this way is the desired Hamiltonian. The eigenvalues $\pm 1$ of
$\sigma_Z^i$ correspond to the positive and negative spin of
(\ref{eq:ising}). After constructing $H$, we can apply VQE to attempt
to determine the ground state, from which an optimal solution to
(\ref{eq:ising}) can be sampled with probability 1.

\begin{figure}
\includegraphics[width=0.45\textwidth]{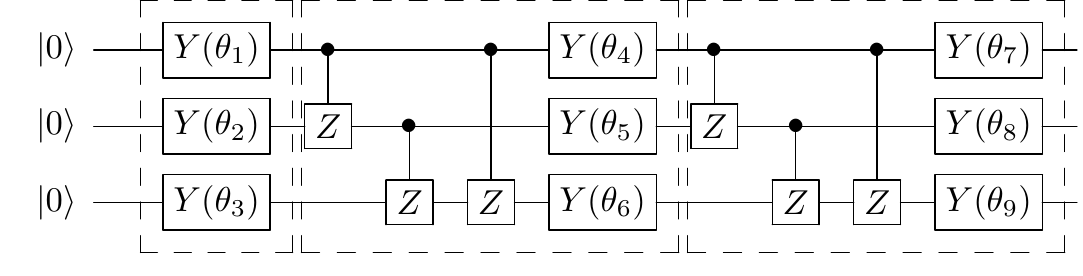}
\caption{VQE variational form for $n=3$ qubits and depth parameter $p=2$. }
\label{fig:vqe_varform}
\end{figure}

The choice of the variational form $U(\theta)$ is important. In this
paper we follow a standard approach, see e.g.,
\cite{kandala2017hardware,Nannicini2019}. Assuming $n$ qubits, we
start by applying single-qubit Y-rotations to every qubit,
parametrized by an angle $\theta_{0,i}$ for qubit $i$. We then repeat
the following $p$ times: we apply controlled Z-gates to all qubit
pairs $(i, j)$ satisfying $i < j$, where $i$ denotes the control qubit
and $j$ the target qubit; and we add another layer of single-qubit Y-rotations to every qubit, parametrized by $\theta_{k,i}$ for qubit $i$
and repetition $k \in \{1, \dots, p\}$. 
Fig.~\ref{fig:vqe_varform} illustrates such a variational form for $n=3$ and $p=2$.
Notice that this variational form span all basis states. Overall, this leads to
$n(1+p)$ parametrized Y-rotations, and $\frac{n(n-1)}{2}p$ controlled
Z-gates.  Since the controlled Z-gates commute with each other, the
total circuit depth is $O(np)$, although the number of gates is
quadratic in $n$. This variational form is used in our numerical simulations. When experimenting on quantum hardware
we use nearest neighbor controlled Z-gates rather than all-to-all, i.e., we follow the connectivity provided by the hardware to
reduce the number of two-qubit gates. We remark that these variational forms span any basis state (and thus the ground state of a diagonal Hamiltonian) already with the first layer of Y-rotations. The purpose of subsequent layers is to hopefully facilitate the task for the classical optimization algorithm used to choose the variational parameters. The above variational form with just one layer of Y-rotations can be efficiently simulated classically; for a discussion on the effect of entangling gates, we refer to \cite{Nannicini2019}.

The classical optimization algorithm that attempts to solve
(\ref{eq:vqe_expected_value}) must be able to (at least) compute the
objective function, i.e., the expected value in
(\ref{eq:vqe_expected_value}). This is difficult to compute directly,
but it can be estimated as follows. First, we prepare the trial
wavefunction $\ket{\psi(\theta)}$ on a quantum processor. Then, we
measure the qubits, resulting in an $n$-bit string $x_0 \dots
x_{n-1}$. Each observed string easily translates into a sample from
$\bra{\psi(\theta)}H\ket{\psi(\theta)}$, because $H$ is a weighted
summation of tensor products of Pauli Z-matrices, and each such term
can be computed with a simple parity check.
We denote these samples by $H_k(\theta)$, $k = 1,\dots,K$, where $K \in \mathbb{N}$ is the number
of samples.  The sample mean
\begin{eqnarray}
\frac{1}{K} \sum_{k=1}^K H_k(\theta) \label{eq:average_objective}
\end{eqnarray}
is an estimator for $\bra{\psi(\theta)}H\ket{\psi(\theta)}$, and is
used as the objective function for the classical optimization
algorithm. The solution returned by the algorithm is then given by the
bitstring that leads to the smallest $H_k$ among all observed
bitstrings and all $\theta$ that have been evaluated.

\section{\label{sec:qaoa} Quantum Approximate Optimization Algorithm}

The Quantum Approximate Optimization Algorithm (QAOA) is a hybrid
quantum/classical algorithm specifically developed for CO problems. In
the context of this paper, QAOA can be seen as a form of VQE with a
specific choice of the variational form that is derived from the
problem Hamiltonian $H$ \citep{Farhi2014}. QAOA enjoys stronger
convergence properties than VQE. For some problems, it can be shown
that QAOA determines a quantum state with a guaranteed approximation
ratio with respect to the ground state
\cite{Farhi2014,farhi2014e3lin2}; furthermore, QAOA applies adiabatic
evolution as the circuit depth goes to $\infty$, implying that it will
determine the optimal solution of the CO problem if the depth of the
variational form is large enough and we can find the optimal circuit
parameters.

Given a problem Hamiltonian corresponding to a QUBO as defined in
(\ref{eq:ising}), the variational form of QAOA is constructed with a
layer of Hadamard gates, followed by two alternating unitaries:
\begin{eqnarray}
U_C(\gamma) &=& e^{-i \gamma \left(\sum_{i=1}^{n} c_i \sigma_Z^i + \sum_{i,j=1}^n Q_{ij} \sigma_Z^i \otimes \sigma_Z^j \right)}, \\
U_B(\beta) &=& e^{-i \beta \sum_{i=1}^{n} \sigma_X^i},
\end{eqnarray}
where $\sigma_X^i$ denotes the Pauli X-matrix
\begin{eqnarray}
\sigma_X^i &=&
\left(
\begin{array}{cc}
0 & 1 \\ 1 & 0
\end{array}
\right) \label{eq:sigma_x}
\end{eqnarray}
acting on the $i$-th qubit.
For a given depth $p \in \mathbb{N}$, the variational form
is thus defined as:
\begin{eqnarray}
U(\beta, \gamma) &=& \left[ \prod_{i=1}^{p} U_B(\beta_i) U_C(\gamma_i) \right] H^{\otimes n},
\end{eqnarray}
where $\beta, \gamma \in \mathbb{R}^{p}$ are vectors of variational
parameters, and here, $H$ denotes a Hadamard gate. (The symbol $H$ is overloaded in the quantum computing literature; in this paper it generally denotes the Hamiltonian, and in the few occasions where it indicates a Hadamard gate, we note it explicitly.)
This yields the trial wave function:
\begin{eqnarray}
\ket{\psi(\beta, \gamma)} &=& U(\beta, \gamma) \ket{0},
\end{eqnarray}
which replaces $\ket{\psi(\theta)}$ in \eqref{eq:vqe_expected_value}.
As in VQE, the expected value is computed as the sample mean over
multiple observations from the quantum state, and the algorithm
returns the bitstring that leads to the smallest $H_k$ among all
observed bitstrings and choices of $\beta, \gamma$.

The number of variational parameters for QAOA is thus $2p$, compared to $n(1+p)$ for VQE.
The circuit depth of the QAOA variational form depends on $n$, $p$, and the number of clauses in the problem
Hamiltonian, which is a difference with respect to VQE.
Every $U_B$ requires $n$ single-qubit X-rotations. One way to construct $U_C$
requires a single-qubit Z-rotation for each $c_i \neq 0$, and a
single-qubit Z-rotation plus two CNOT-gates for each $Q_{ij} \neq 0$,
yielding the term $e^{-i\gamma \sigma_Z^i \otimes \sigma_Z^j}$ up to a
global phase.  The circuits used in our implementation are described
in Appendix \ref{sec:qaoa_circuit}.  In total, this leads to $O(n^2p)$
single-qubit rotations and $O(n^2p)$ CNOT-gates, but the scaling may
be better than the worst case if $Q$ is sparse (i.e., there are few
two-qubit interactions).

\section{\label{sec:cvar} CVaR Optimization}

In quantum mechanics observables are defined as expected values
$\bra{\psi}H\ket{\psi}$. This leads to the natural choice of the
sample mean (\ref{eq:average_objective}) as the objective function for the
classical optimization problems embedded in VQE and QAOA. We argue
that for problems with a diagonal Hamiltonian, such as CO problems,
the sample mean may be a poor choice in practice. This is because when
$H$ is diagonal, there exists a ground state which is a basis
state. If determining the value $H_{j,j}$ of a basis state $\ket{j}$
is classically easy, the state with the minimum eigenvalue computed by
an algorithm is simply the best measurement outcome among all
measurements performed. It is therefore reasonable to focus on
improving the properties of the best measurement outcome, rather than
the average. We illustrate this intuition with a simple
example. Consider two algorithms ${\cal A}_1$ and ${\cal A}_2$ applied
to a problem with Hamiltonian $H$, minimum eigenvalue $\lambda_{\min}$
and ground state $\ket{j}$. Suppose ${\cal A}_1$ produces a state
$\ket{\psi_1}$ and ${\cal A}_2$ produces a state $\ket{\psi_2}$, with
the following properties: $\bra{\psi_1}H\ket{\psi_1} / \lambda_{\min}
= 1.1$ and $\braket{j|\psi_1} = 0$; $\bra{\psi_2}H\ket{\psi_2} /
\lambda_{\min} = 2.0$ and $\braket{j|\psi_2} = 0.1$. We argue that
from a practical point of view, algorithm ${\cal A}_2$ is likely to be
more useful than ${\cal A}_1$. This is because even if ${\cal A}_1$
leads to samples with a better objective function value on average,
${\cal A}_1$ will never yield the ground state $\ket{j}$; whereas
${\cal A}_2$, which is much worse on average, has a positive and
sufficiently high probability of yielding the ground state, so that
with enough repetitions we can be almost certain of determining
$\ket{j}$.

In light of this discussion, one way to attain our goal would be to
choose, as the objective function, the minimum observed outcome over a set
of measurements: $\min \{ H_1, ..., H_K \}$.
However, for finite $K$ this typically leads to a non-smooth,
ill-behaved objective function that is difficult to handle for
classical optimization algorithms.

To help smooth the objective function, while still focusing on
improving the best measured outcomes rather than the average, we
propose the Conditional Value at Risk (CVaR, also called Expected
Shortfall) as the objective function.  Formally, the CVaR of a random variable
$X$ for a confidence level $\alpha \in (0, 1]$ is defined as
\begin{eqnarray}
\text{CVaR}_{\alpha}(X) &=& \mathbb{E}[X | X \leq F_X^{-1}(\alpha)]
\end{eqnarray}
where $F_X$ denotes the cumulative density function of $X$. In other
words, CVaR is the expected value of the lower $\alpha$-tail of the
distribution of $X$. Without loss of generality, assume that the
samples $H_k$ are sorted in nondecreasing order, i.e. $H_{k+1} \geq
H_k$. Then, for a given set of samples $\{H_k\}_{k=1,\dots,K}$ and
value of $\alpha$, the $\text{CVaR}_{\alpha}$ is defined as
\begin{eqnarray}
\frac{1}{\lceil \alpha K \rceil} \sum_{k=0}^{\lceil \alpha K \rceil} H_k.
\end{eqnarray}
Note that the limit $\alpha \searrow 0$ corresponds to the minimum, and $\alpha = 1$
corresponds to the expected value of $X$. In this sense, CVaR is a
generalization of both the sample mean (\ref{eq:average_objective}),
and the best observed sample $\min \{ H_1, ..., H_K \}$. For small,
nonzero values of $\alpha$, CVaR still puts emphasis on the best
observed samples, but it leads to a smoother and easier to handle
objective function. It is clear that this can be applied to both VQE
and QAOA, simply by replacing the sample mean
(\ref{eq:average_objective}) with $\text{CVaR}_{\alpha}$ in the
classical optimization algorithm. We call the resulting algorithms
\cvar-VQE and \cvar-QAOA,
respectively.

\section{\label{sec:cvar_analysis} Analysis of CVaR Optimization}

The optimization of $\cvar_{\alpha}$ with $\alpha < 1$ modifies the
landscape of the objective function of VQE and QAOA as compared to the
expected value, i.e., $\alpha = 1$. This is formalized next.

\begin{figure}
  \leavevmode
  \centering
  \Qcircuit @C=1em @R=.6em {
    & \qw & \gate{H} & \ctrl{1} & \qw & \qw\\
    & \qw & \qw      & \targ    & \gate{R_{Y}(\theta)} & \qw\\
  }
  \caption{Variational form for the first part of
    Prop.~\ref{prop:local_min_mapping}. Here, $H$ indicates a
    Hadamard gate.}
  \label{fig:var_form_ex1}
\end{figure}
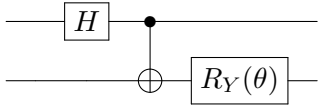

We need to define a random variable that encodes the classical
objective function value of a measurement outcome, i.e., the value of
a binary string in the QUBO problem. Let $X(\theta)$ be the
random variable with outcomes $H_{j,j}$ for $j \in
\{0,1\}^n$, i.e., the diagonal elements of the Hamiltonian, and
$\text{Prob}(X(\theta) = H_{j,j}) = |\alpha_{j}(\theta)|^2$ where
$\ket{\psi(\theta)} = \sum_{j} \alpha_j(\theta) \ket{j}$. In other
words, $X(\theta)$ represents the QUBO objective function associated
with a single measurement taken on the quantum state
$\ket{\psi(\theta)}$. Then the \cvar-version of
(\ref{eq:vqe_expected_value}) can be written as:
\begin{eqnarray}
  \min_{\theta} \cvar_\alpha(X(\theta)) \label{eq:vqe_cvar}.
\end{eqnarray}
As it turns out, there is no well-defined mapping between local minima
of (\ref{eq:vqe_expected_value}) and (\ref{eq:vqe_cvar}).
\begin{proposition}
  \label{prop:local_min_mapping}
  A local minimum of (\ref{eq:vqe_expected_value}) does not
  necessarily correspond to a local minimum of (\ref{eq:vqe_cvar}),
  and vice versa.
\end{proposition}
To show this, we first exhibit a local minimum of
(\ref{eq:vqe_expected_value}) that is not a local minimum of
(\ref{eq:vqe_cvar}). Consider a two-qubit Hamiltonian $H =
\text{diag}(0,1,1,2)$, and variational form depicted in
Fig.~\ref{fig:var_form_ex1}.
It is easy to verify that the quantum state constructed by this
variational form is $\ket{\psi(\theta)} = \frac{1}{\sqrt{2}} (\cos
\frac{\theta}{2}, \sin \frac{\theta}{2}, -\sin \frac{\theta}{2}, \cos
\frac{\theta}{2})$. Given the Hamiltonian, we have $\cvar_1(X(\theta))
= \mathbb{E}[X(\theta)] = \cos^2\frac{\theta}{2} + \sin^2 \frac{\theta}{2} = 1$ independent of
$\theta$, so every value of $\theta$ is a local (in fact, global)
minimum. On the other hand, it is clear that for $\alpha < 1$ there
are values of $\theta$ that are not local minima: for example, with
$\alpha = 0.5$ doing the calculations shows that
$\cvar_{0.5}(X(\theta)) = \sin^2 \frac{\theta}{2}$. Hence,
the only local minima are at $\theta = 2k\pi$ for $k$ integer.

The converse, i.e., a local minimum of (\ref{eq:vqe_cvar}) that is not
a local minimum of (\ref{eq:vqe_expected_value}), is trivial: for any
problem (\ref{eq:vqe_expected_value}) and parameters $\theta^*$ such
that $\ket{\psi(\theta^*)}$ has overlap $\rho > 0$ with the ground
state, $\theta^*$ is a global minimum of $\cvar_{\alpha}(X(\theta^*))$
for $\alpha \le \rho$, even if $\theta^*$ is not a local minimum of
(\ref{eq:vqe_expected_value}).

It is easy to see from the above discussion on local minima that, in
fact, we cannot even map global minima of problems (\ref{eq:vqe_expected_value})
and (\ref{eq:vqe_cvar}) to each other. In fact, suppose a
certain trial state $\ket{\psi(\theta)}$ has overlap $\rho$ with the
ground state, then it is a global optimum of (\ref{eq:vqe_cvar}) for
any $\alpha \le \rho$.
On the other hand, it is clear that it may not
be a global optimum of (\ref{eq:vqe_expected_value}), depending on the variational form chosen. 
However, a specific case of interest for this paper is that, in which the variational form is capable of
reaching the ground state, and the Hamiltonian is diagonal; then the above discussion implies that a global
optimum of Eq.(\ref{eq:vqe_expected_value}) is also a
global optimum of Eq.(\ref{eq:vqe_cvar}) for any $\alpha$.

\begin{figure}
 \centering
 \includegraphics[width=0.4\textwidth]{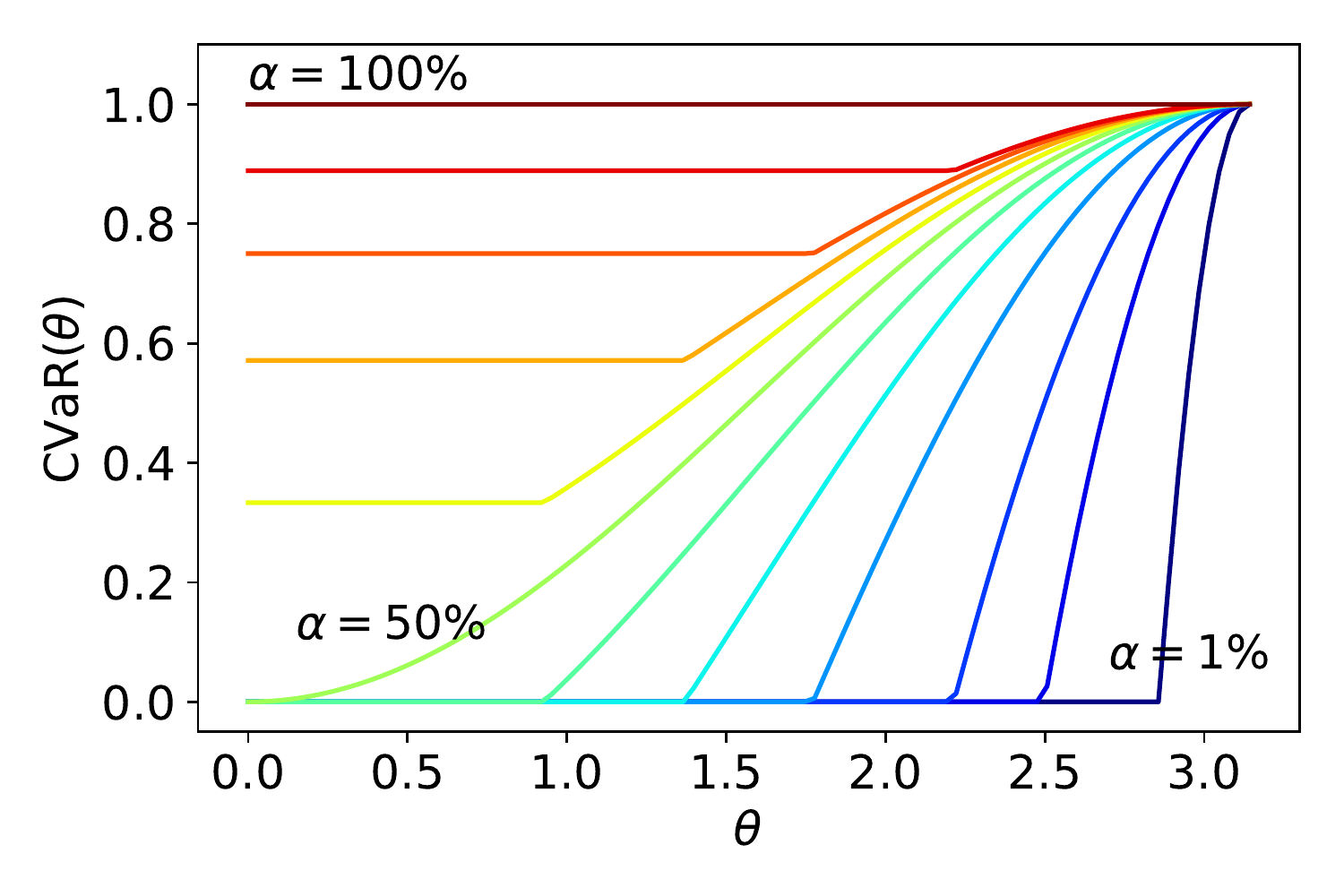}
 \caption{Objective function for different values of $\theta$ and
   $\alpha$: the top line corresponds to $\alpha = 1$, the remaining
   lines show decreasing values of $\alpha$ down to $\alpha = 0.01$
   for the line in the bottom right. }
\label{fig:cvar_2q_example}
\end{figure}

Even if Prop.~\ref{prop:local_min_mapping} indicates that mapping
properties of the optimization problems (\ref{eq:vqe_expected_value})
and (\ref{eq:vqe_cvar}) is not trivial, the two-qubit example
exhibiting a local minimum of (\ref{eq:vqe_expected_value}) that is
not a local minimum of (\ref{eq:vqe_cvar}) showcases a situation in
which \cvar~optimization is clearly preferrable. Indeed, using
(\ref{eq:vqe_expected_value}) yields a constant objective function on
which no optimization can be performed. However, \cvar~with $0 <
\alpha < 1$ yields a smooth objective function that is optimized by
decreasing $\theta$. This has a positive effect on the probability of
sampling the ground state $\ket{00}$, which is maximized at $\theta =
0$ in the interval $[0, \pi]$. The objective function
(\ref{eq:vqe_cvar}) in this example for different values of $\alpha$
is illustrated in Fig.~\ref{fig:cvar_2q_example}. 

It is important to remark that the natural empirical estimator of CVaR$_{\alpha}$ considers a subset of the measurements only.
This raises the question of how to choose the number of samples for a particular value of $\alpha$ to achieve a certain accuracy.
The variance of the empirical CVaR$_{\alpha}$ estimator using $K$ samples is $\mathcal{O}(1 / (K \alpha^ 2) )$ (see e.g., \cite{hong2011mc_var_cvar}), implying that the resulting standard error increases as $1 / \alpha$.
Thus, for a fixed number of samples $K$, to achieve the same accuracy as for the expected value we need to increase the number of samples to $K / \alpha$.
(We remark that although \cite{hong2011mc_var_cvar} only shows this dependency for continuous distributions, we expect it to be a good approximation as the number of qubits increases.) As our numerical experiments show, $\alpha$ can be chosen as a constant, independent of the number of qubits, resulting in a constant increase of the number of samples for fixed accuracy. A discussion on the accuracy of the estimation from an empirical point of view is given in Sec.~\ref{sec:experimental_setup}.

\section{\label{sec:experimental_setup} Computational experiments}

The preceding analysis shows that \cvar~optimization may improve
certain properties of the classical optimization problem solved in VQE
and QAOA. To verify if this is the case from an empirical point of
view, we test the proposed on multiple random instances of six CO
problems: maximum stable set, maximum 3-satisfiability, number
partitioning, maximum cut, market split, and portfolio
optimization. Below, we give a brief description of these problems. A
more detailed discussion of the instance generation and the mapping to
a Hamiltonian can be found in \cite{Nannicini2019} for all problems
except portfolio optimization, which is discussed in Appendix 
\ref{sec:portfolio_optimization_details}.

\begin{description}

\item[Maximum stable set] Given an undirected graph $G = (V, E)$, a
  stable set (also called independent set) is a set of mutually
  nonadjacent vertices. The objective of the maximum stable set
  problem is to find a stable set of maximum
  cardinality.

\item[Maximum 3-satisfiability] The objective of the maximum
  3-satisfiability problem is to find an assignment
  of boolean variables that satisfies the largest number of clauses of
  a boolean formula in conjunctive normal form, where each clause has
  exactly three literals.

\item[Number partitioning] Given a set of numbers $S = \{a_1, ...,
  a_n\}$, the problem of number partitioning asks
  to determine disjoint sets $P_1, P_2 \subset \{1, ..., n\}$ with
  $P_1 \cup P_2 = \{1, ..., n\}$, such that $\left| \sum_{i \in P_1}
  a_i - \sum_{j \in P_2} a_j \right|$ is minimized.

\item[Maximum cut] Given a weighted undirected graph $G = (V, E)$
  with edge weights $w_{ij}$, the maximum cut problem
  aims to determine a partition of $V$ into two
  disjoint sets $V_1, V_2$ such that the sum of weights of edges that
  connect $V_1$ and $V_2$ is maximized.

\item[Market split] The market split problem can be described as the 	
  problem of assigning $n$ customers of a firm
  that sells $m$ products to two subdivisions of the same firm, such
  that the two subdivisions retain roughly an equal share of the
  market.

\item[Portfolio optimization] Given a set of $n$ assets $\{1, ...,
  n\}$, corresponding expected returns $\mu_i$ and covariances
  $\sigma_{ij}$, a risk factor $q > 0$ and a budget $B \in \{1, ...,
  n\}$, the considered portfolio optimization problem
  tries to find a subset of assets $P \subset
  \{1, ..., n\}$ with $|P| = B$ such that the resulting $q$-weighted
  mean-variance, i.e. $\sum_{i \in P} \mu_i - q\sum_{i,j \in P}
  \sigma_{ij}$, is maximized.

\end{description}

For each problem except \textsc{Max3Sat}, we generate ten random
instances on $6, 8, 10, 12, 14$, and $16$ qubits. Our formulation of
\textsc{Max3Sat} requires the number of qubits to be a multiple of
three, thus we use $6, 9, 12$, and $15$ qubits. For every instance, we
run \cvar-VQE and \cvar-QAOA for $\alpha \in \{1\%, 5\%, 10\%, 25\%,
50\%, 75\%, 100\%\}$ and $p = 0, 1, 2$ for VQE and $p = 1, 2, 3$ for
QAOA. In total, this leads to 340 random problem instances and
$14,280$ test cases. Following \cite{Nannicini2019}, we use the
classical optimizer COBYLA to determine the parameters of the trial
wave function.

In the first part of our experimental evaluation
(Sec.~\ref{sec:simulation_results}), we analyze the performance of the
different algorithms using the exact quantum state resulting from
simulation. This allows us to precisely characterize the performance
metrics that we use. In the second part
(Sec.~\ref{sec:quantum_results}), we study the performance of the proposed
approach using existing quantum hardware.

\subsection{\label{sec:simulation_results} Results I: Simulation}

We compare the different algorithms by plotting the resulting
probability of sampling an optimal solution versus the number of
iterations of the classical optimization algorithm. To make the number
of iterations comparable for problems of different sizes, we normalize
it dividing by the number of qubits. We choose the probability of
sampling an optimal solution, rather than some aggregate measure of
the objective function value, because all our algorithms use different
metrics in this respect: comparing algorithms with respect to the
average objective function value (or \cvar~with a different $\alpha$)
would not be informative. The probability of sampling an optimal
solution (i.e., the overlap with ground state) is a reasonable metric
that provides valuable information across different values of
$\alpha$.

\begin{figure*}[ht!]
\centering
\includegraphics[width=0.8\textwidth]{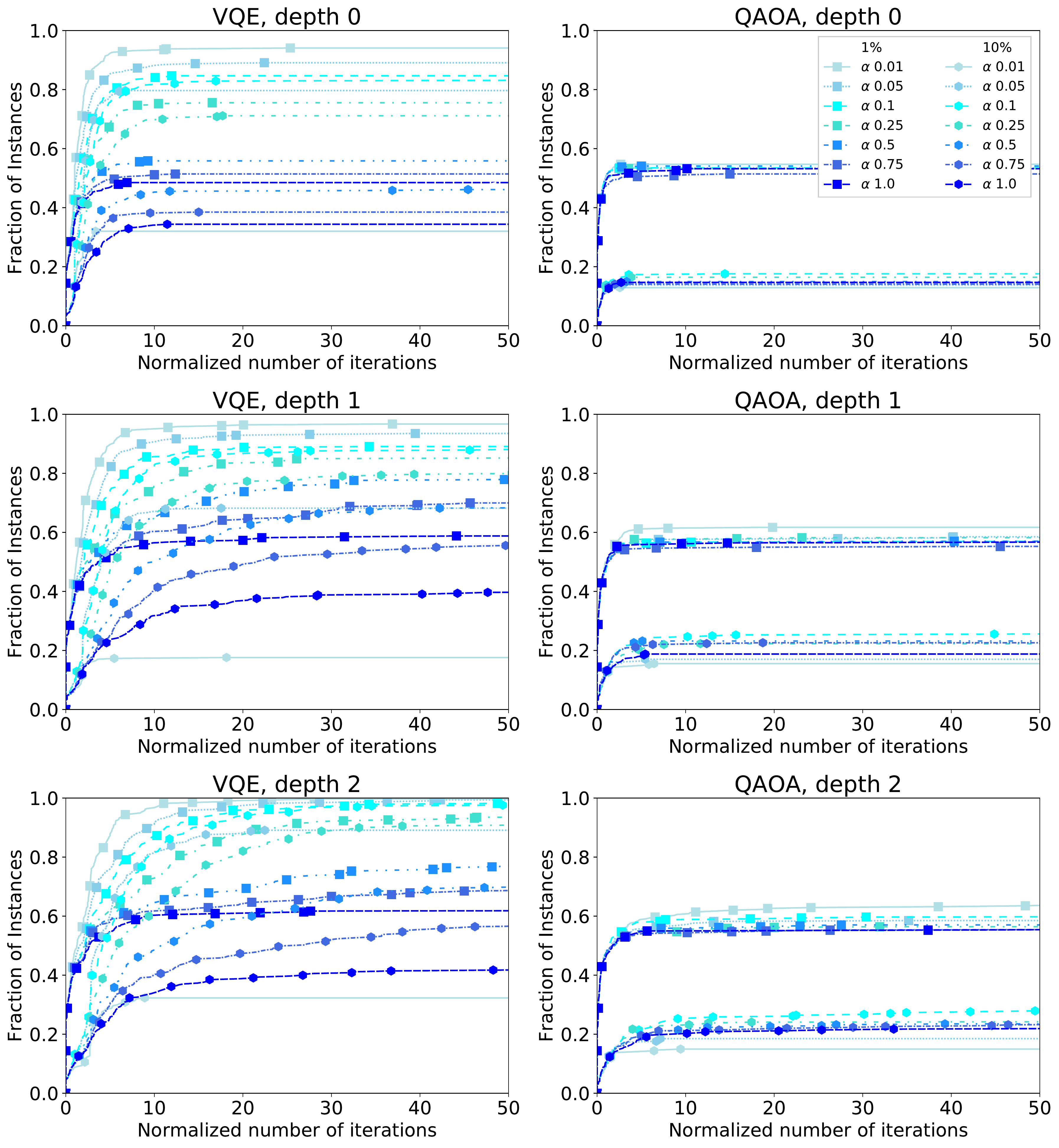}

\caption{
Summary of the results with VQE (left) / QAOA (right) using quantum states resulting from classical simulation.
On the $x$-axis we plot the normalized number of iterations, on the $y$-axis the fraction of the instances that attain a certain probability of sampling an optimal basis state.
Each plot contains results for different levels of $\alpha$ (reported as $aX\%$ in the legend), and the depth increases from top to bottom.}
\label{fig:vqe_qaoa_sample_profiles}

\end{figure*}

Fig.~\ref{fig:vqe_qaoa_sample_profiles} shows the fraction of instances
that achieve at least a certain probability of sampling an optimal
state ($\geq 1\%$ and $\geq 10\%$) with respect to the number of
normalized iterations for CVaR-VQE / CVaR-QAOA (using all-to-all entanglement).
This fraction is shown for each considered variational form depth, $p$, and
value of $\alpha$.  The plots show that increasing $p$ and decreasing
$\alpha$ has a positive impact on the performance. 

For CVaR-VQE, using $p=2$ and
$\alpha=1\%$, within 50 normalized iterations we achieve at least
$1\%$ probability of sampling an optimal state for almost all
instances. In contrast, with $\alpha = 100\%$ (i.e., the expected
value), we reach the same probability of sampling an optimum only for
$60\%$ of the test problems. Notice that the value of $\alpha$
introduces a soft cap on the maximum probability of sampling a ground
state: for example, for $\alpha = 10\%$ we reach $10\%$ probability to
sample an optimal solution for most of the test problems in less than
50 normalized iterations, but with $\alpha=1\%$ we reach $10\%$
probability only in a small fraction of problems. This is expected,
because the \cvar~objective function with $\alpha = 1\%$ does not
reward increasing the overlap with the ground state beyond 1\%
probability.

For CVaR-QAOA, we observe improved performance as $p$ increases and $\alpha$ decreases
(up to a certain level). Comparing VQE and QAOA, we observe that QAOA's performance
appears significantly worse than that of VQE for equivalent depth (where we compare depth $p$ for VQE to depth $p+1$ for QAOA).  We
conjecture that this is due to the limited number of variational
parameters in QAOA: only $2p$, compared to $n(1+p)$ for VQE. An
intuitive explanation is that the state vector obtained with QAOA is
thus relatively ``flat'', and never reaches a large overlap with the
ground state. This intuitive explanation is formalized in
Sec.~\ref{sec:qaoa_analysis}. In the context of this paper, one of
QAOA's characteristics, i.e., the concentration around the mean
\cite{Farhi2014}, may become a weakness in the practical context of
sampling the optimum (or a near-optimal solution) with sufficiently
large probability. To improve QAOA's performance we would have to
increase the depth. Since the current generation of quantum hardware
is affected by non-negligible gate errors and decoherence, successfully
implementing circuits with large depth may be out of reach for the
moment.

\begin{figure*}[ht!]
\centering
\includegraphics[width=0.8\textwidth]{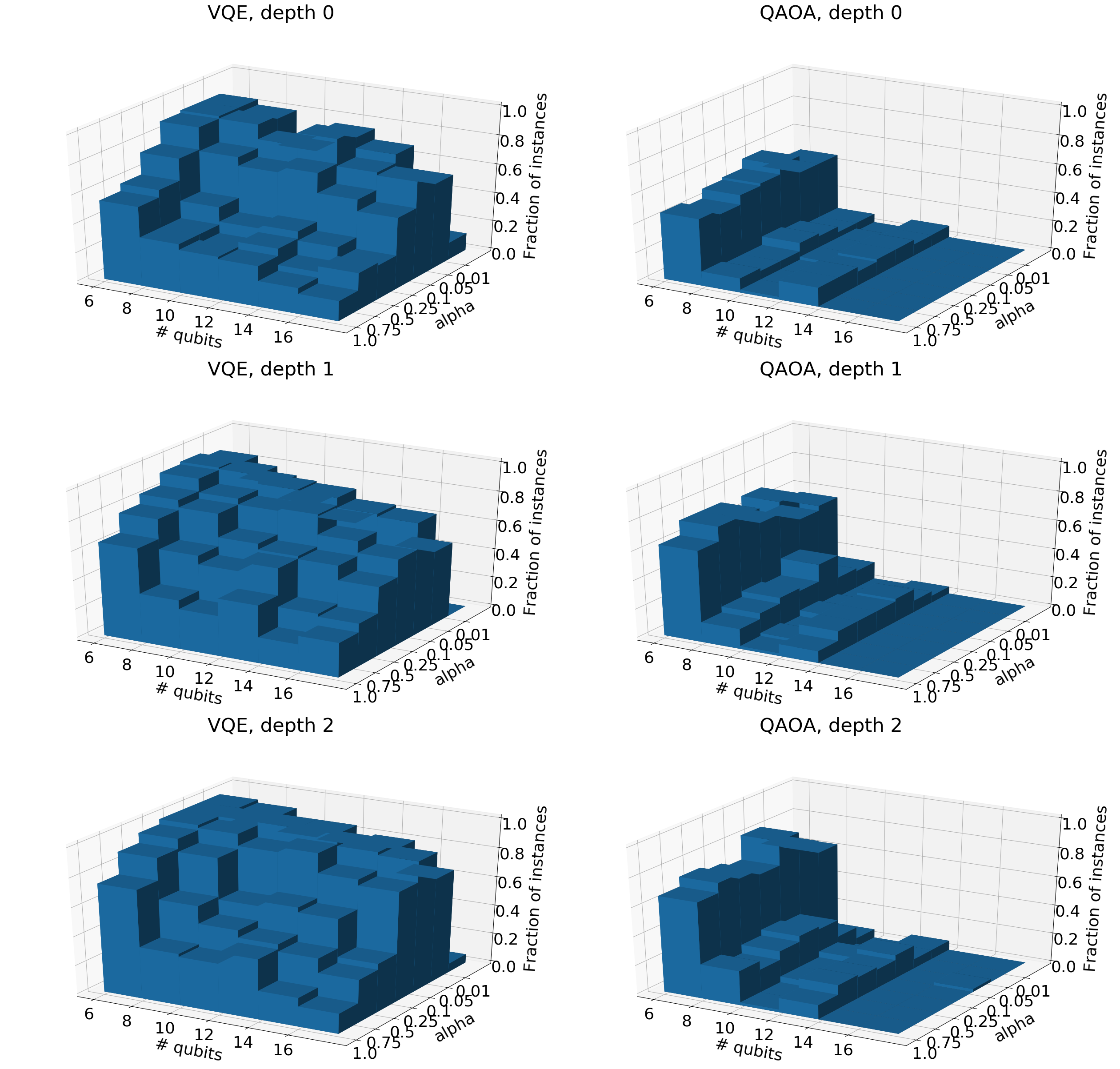}

\caption{
Summary of the results with VQE (left) / QAOA (right) using quantum states resulting from classical simulation, after 50 normalized iterations of the classical optimization algorithm. On the vertical $z$-axis we plot the fraction of instances that attain $10\%$ probability of sampling the optimal solution, on the $xy$-plane we indicate the number of qubits and the value of $\alpha$.}
\label{fig:vqe_qaoa_3d_prob_01}

\end{figure*}

To ensure that the positive effect of CVaR optimization is not lost
when the problem size increases, we look at the results across
different number of qubits and values of $\alpha$. The corresponding
plots are given in Fig.~\ref{fig:vqe_qaoa_3d_prob_01} for a
probability of sampling the optimum of 10\%; a similar figure for 1\% probability is available in Appendix~\ref{sec:additional_plots}. 
Fig.~\ref{fig:vqe_qaoa_3d_prob_01} shows that for a small number of qubits there is a ceiling effect,
i.e., all methods perform similarly because the problem is easy for
all methods, but as soon as problem size increases, the benefits of
CVaR optimization (with $\alpha \in [0.01, 0.25]$) are obvious in the
plots.

\subsection{\label{sec:quantum_results} Results II: Quantum Device}

To test CVaR optimization on quantum hardware, we consider an instance
of the portfolio optimization problem with 6 assets mapped to 6
qubits, see Appendix \ref{sec:portfolio_optimization_details}. We
choose portfolio optimization because the problems of this class are
some of the most difficult of our testbed.

We test CVaR-VQE on the \emph{IBM Q Poughkeepsie} 20-qubit quantum
computer, with COBYLA as the classical optimizer.  In this section we
apply nearest neighbor entanglement instead of all-to-all
entanglement.  We choose 6 qubits on the device that are connected in
a ring (qubits 5, 6, 7, 10, 11, and 12), thus achieving a cyclic
entanglement without additional swap operations; see Appendix
\ref{sec:portfolio_optimization_details} for more detail.  We use
CVaR-VQE rather than CVaR-QAOA because for the same circuit depth it
leads to better solutions, as discussed in
Sec.~\ref{sec:simulation_results}.

We run CVaR-VQE with depth $p = 1$ and $\alpha = 10\%, 25\%, 100\%$,
repeating each experiment five times. We gather 8,192 samples from
each trial wavefunction, studying the probability of measuring a
ground state with respect to the number of iterations of the classical
optimization algorithm. To illustrate the convergence of the
algorithm, we also plot the progress in the objective function value.
Note that the reported objective function values for different values
of $\alpha$ are incomparable. To reduce variance in the experiments,
we fix the initial variational parameters to $\theta = 0$. Results are
reported in Fig.~\ref{fig:real_hardware_results}. Similar plots for
depth $p=0$ and $p=2$ are given in Appendix
\ref{sec:portfolio_optimization_details}.

We see that the smaller the $\alpha$, the earlier the probability of
sampling an optimal solution increases.  For $\alpha = 100\%$, the
probability stays almost flat and makes little progress.  The plots of
the objective function also show that $\alpha < 100\%$ speeds up the
convergence of the objective function values to a (local) optimum.
For $\alpha = 10\%, 25\%$ the probability of finding the optimal
solution attains the corresponding $\alpha$-level in all 5
experiments, whereas for $\alpha = 100\%$ the probability remains very
small. Recall that the CVaR objective function does not provide any incentive to increase the overlap with the optimal solution beyond $\alpha$.

\begin{figure}[ht]
\includegraphics[width=0.5\textwidth]{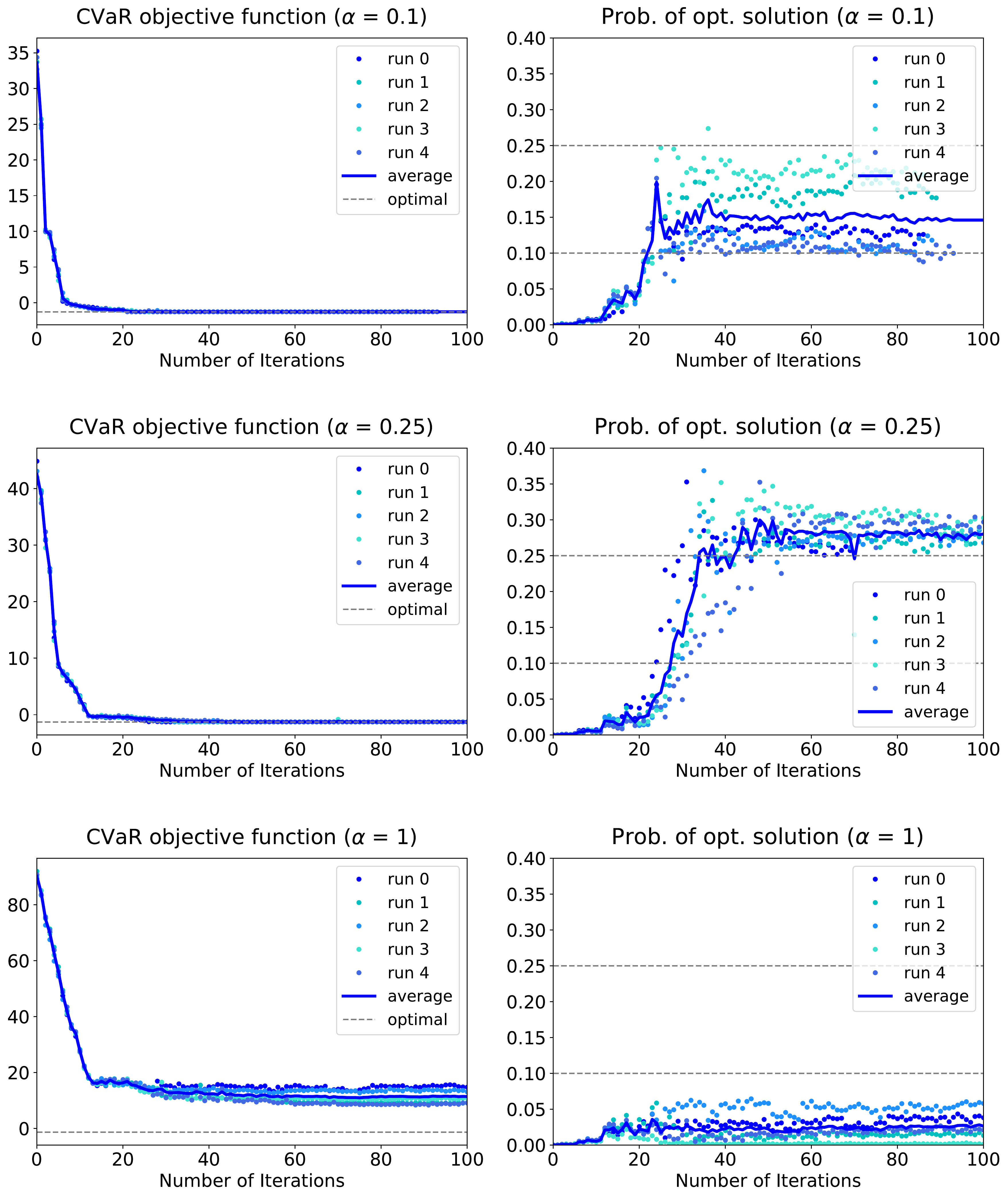}
\caption{Quantum hardware results for portfolio optimization problem
  with six assets/qubits. CVaR-VQE results are shown for depth $p=1$ and
  $\alpha=10\%, 25\%, 100\%$ (from top to bottom) and five runs with
  8,192 samples for each $\alpha$. Plots on the left: resulting objective
  values per iteration; plots on the right: resulting probability of
  sampling an optimal solution.  Since COBYLA converges after a
  different number of iterations in each run, we assume that the contribution of each run to the average value after termination of that run is its last reported value.
  The $\alpha$-levels $10\%$ and $25\%$ are indicated by the
  gray dashed lines in all probability plots.}
\label{fig:real_hardware_results}
\end{figure}

In addition to the improved convergence behavior already demonstrated
in Sec.~\ref{sec:simulation_results} using classical simulation, the
CVaR objective function also seems to be able to cope with the noise
and errors introduced by the quantum hardware. Indeed, on quantum
hardware we observe the same beneficial effect on the speed of
convergence that was observed in the noiseless simulation results.
A possible explanation is that the CVaR objective function allows us to ignore some of the low-quality samples from the quantum state. In
other words, even if we do not reach the ground state (which may be
difficult to detect in the presence of noise), CVaR focuses on
ensuring that at least {\em some} of the samples have a low objective
function value, which may be a more attainable goal and seems to drive the classical optimization algorithm in the right direction. This effect makes the CVaR objective particularly well-suited for
experiments on noisy quantum computers.

We end this section with a discussion on the impact of $\alpha$ on the number of samples. As discussed in Sec.~\ref{sec:cvar_analysis}, to obtain the same accuracy as the expected value we would need to increase the number of samples by a factor $\frac{1}{\alpha}$. Results in this paper suggest choosing $\alpha \in [0.1, 0.25]$ as a good empirical choice, implying that the number of samples should be increased by a factor $[4, 10]$ to attain the same accuracy. However, our empirical evaluation uses the same, fixed number of samples across all $\alpha$, and still shows significant benefits of \cvar~optimization. A possible explanation is that as long as the number of samples allows a reasonable estimation of the \cvar~objective function, the loss in estimation accuracy (as compared to the expected value) is counterbalanced by the fact that the \cvar~objective is more effective at guiding the classical optimization algorithm toward a quantum state that overlaps with the optimal solution. Thus, in our empirical evaluation even a noisy \cvar~estimate yields better results than a more accurate expected value estimate.

\section{\label{sec:qaoa_analysis} On the performance of QAOA} 
In this section we formalize our intuition that, due to the small
number of variational parameters, QAOA may produce relatively ``flat''
state vectors, i.e., with amplitudes of similar magnitude. We
initially observed this behavior empirically, and it can be made
precise under some additional conditions.
\begin{proposition}
  \label{prop:qaoa}
  Assume that we apply QAOA to a problem on $n$ qubits with objective
  function encoded in a diagonal Hamiltonian $H$ with diagonal
  elements $H_{j,j}, j=1,\dots,2^n$. 
  Let $\delta := \max_{\ell} |\{ j \in \{0,1\}^n : H_{j,j} = \ell\}| / 2^n$, i.e., the maximum fraction
  of basis states having the same eigenvalue. Let $\ket{\psi_p} =
  \sum_{z} \alpha_{p,z} \ket{z}$ be the quantum state produced by QAOA
  with depth $p$.  Let $\Delta_p$ be a lower bound on the maximum
  fraction of amplitudes that are equal after iteration $p$ of QAOA,
  i.e., $\Delta_p := \min_{t \le p} \max_{u \in \mathbb{C}} |\{j \in
  \{0,1\}^n: \alpha_{t,j} = u \}|/2^n$. Then $|\alpha_{p,z}| \le
  (2^{n+1} (2 - \Delta_{p-1} -\delta ) + 1)^{p} \frac{1}{\sqrt{2^n}}$
  for all $z \in \{0,1\}^n$. Furthermore, $\Delta_0 = 1$, but the
  value of $\Delta_p$ may decrease exponentially fast in $p$.
\end{proposition}
The proof of the above proposition is given in Appendix
\ref{sec:proof_qaoa}. While the statement is technical, we discuss
some special cases that provide an intuition. When $p = 1$ and most of
the diagonal values of the Hamiltonian are equal, say, $\delta \ge 1 -
2^{-n(\frac{1}{2} + \epsilon)}$, $\epsilon > 0$, then the resulting state vector is
necessarily flat: all the amplitudes are exponentially small
$O(\frac{1}{2^{\epsilon n}})$. This situation is easy to envision:
when the Hamiltonian does not provide enough information on the
distribution of objective function values, QAOA with small depth
cannot transfer enough probability mass to any basis state. This is
the case, for instance, in the Grover ``needle in a haystack'' problem
\cite{grover1997quantum} where a unique $z^*$ has objective function
value $H_{z^*,z^*} = 1$ and $H_{z,z} = 0$ for all other $z \in
\{0,1\}^n$. Another example is given by the feasibility version of the
market split problems, see the analysis on the number of solutions in
\cite{aardal2000market}. The Hamiltonians that necessarily lead to
flat state vectors are those with $\delta, \Delta_p \ge 1 -
2^{-n(\frac{1}{2} + \epsilon)}$; notice that intuitively, $\delta
\approx 1$ is more likely to lead to $\Delta_p \approx 1$, although
our proof in Appendix \ref{sec:proof_qaoa} shows a very loose lower
bound on $\Delta_p$ that is exponentially decreasing in $p$. Another
way to interpret Prop.~\ref{prop:qaoa} is that QAOA requires the
diagonal elements of the Hamiltonian (i.e., objective function values)
to be well-distributed to effectively ``mix'' and increase the
amplitudes. This can also be achieved increasing $p$, say, linearly in
$n$, but for fixed $p$ there is the risk that the amplitudes remain
flat. While this may still lead to a good average objective function
value, it may not put enough emphasis on the tail of the distribution
of objective function values to sample an optimal or a near-optimal
solution. Notice that examples of this are also discussed in the
seminal paper \cite{Farhi2014}: the paper shows that for MaxCut on
2-regular graphs, QAOA produces a state with approximation ratio $3/4$
but exponentially small overlap with the optimal solution.

\section{\label{sec:conclusions} Conclusions}

We introduce improved versions of the hybrid quantum/classical
algorithms VQE and QAOA for CO, based on the CVaR aggregation function for the samples obtained from trial wavefunctions.
We provide theoretical and empirical results, showing an increase in performance compared to approaches in the literature.
This includes a demonstration on IBM's quantum hardware, where the algorithm that we propose shows the ability to reach an optimal solution much faster.

\subsection*{Code Availability}
A notebook providing the code to run CVaR-VQE is available open source at
\url{https://github.com/stefan-woerner/cvar_quantum_optimization/}

\begin{acknowledgments}
S.~Woerner is partially supported by AFRL grant
FA8750-C-18-0098. G.~Nannicini is partially supported by AFRL grant
FA8750-C-18-0098 and the Research Frontiers Institute.

IBM, IBM Q, Qiskit are trademarks of International
Business Machines Corporation, registered in many ju-
risdictions worldwide. Other product or service names
may be trademarks or service marks of IBM or other
companies.
\end{acknowledgments}



\appendix

\section{\label{sec:qaoa_circuit} Implementation of QAOA}
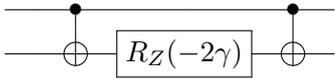
\begin{figure}[ht]
  \leavevmode
  \centering
  \Qcircuit @C=1em @R=.6em {
    & \qw & \ctrl{1} & \qw & \ctrl{1} & \qw\\
    & \qw & \targ    & \gate{R_Z(-2\gamma)} & \targ & \qw \\
  }
  \caption{Implementation of $e^{-i\gamma \sigma_Z^i \otimes \sigma_Z^j}$ up to a global phase.}
  \label{fig:qaoa_zz}
\end{figure}

As discussed in the main text, implementing the blocks $U_B$ and $U_C$
of QAOA requires both single-qubit rotations and CNOT-gates.
The implementation of $U_B$ is trivial in Qiskit, as X-rotations $R_X$ are natively supported, we can apply $R_X(-2\beta)$ to each qubit to implement $U_B$.
For $U_C$, a possible implementation for $e^{-i\gamma\sigma_Z^i \otimes \sigma_Z^j}$ up to a global phase is depicted in Fig.~\ref{fig:qaoa_zz} using two CNOT-gates and one single-qubit Z-rotation $R_Z$ (both also natively supported in Qiskit).

\section{\label{sec:portfolio_optimization_details} Portfolio Optimization on Quantum Device}

The portfolio optimization problem considered in Sec.~\ref{sec:quantum_results} is given by
\begin{eqnarray*}
\max_{x \in \{0,1\}^n} \sum_{i=1}^{n} \mu_i x_i - q \sum_{i,j=1}^n \sigma_{ij} x_i x_j - \lambda \left(B - \sum_{i=1}^n x_i \right)^2,
\end{eqnarray*}
where we subtract a penalty term weighted by $\lambda$ to enforce the budget constraint $\sum_{i=1}^n x_i = B$.
We choose $n = 6$, $q = 0.5$, $B = 3$, and $\lambda = 12$.
The used return vector $\mu$ and positive semidefinite covariance matrix $\sigma$ were generated randomly and are given by:
\begin{eqnarray*}
&&\tiny
\mu 
\tiny
=
\left(
\begin{array}{cccccc}
0.7313 & 0.9893 & 0.2725 & 0.8750 & 0.7667 & 0.3622
\end{array}
\right) \\
&&\tiny
\sigma
\tiny
=\\
&&\tiny
\left(
\begin{array}{rrrrrr}
	 0.7312 & -0.6233 &  0.4689 & -0.5452 & -0.0082 & -0.3809 \\
	-0.6233 &  2.4732 & -0.7538 &  2.4659 & -0.0733 &  0.8945 \\
	 0.4689 & -0.7538 &  1.1543 & -1.4095 &  0.0007 & -0.4301 \\
	-0.5452 &  2.4659 & -1.4095 &  3.5067 &  0.2012 &  1.0922 \\
	-0.0082 & -0.0733 &  0.0007 &  0.2012 &  0.6231 &  0.1509 \\
	-0.3809 &  0.8945 & -0.4301 &  1.0922 &  0.1509 &  0.8992
\end{array}
\right)
\end{eqnarray*}
The corresponding Hamiltonian can be constructed as described e.g.~in Sec.~\ref{sec:vqe} and the references mentioned therein.

\begin{figure}
\includegraphics[width=0.5\textwidth]{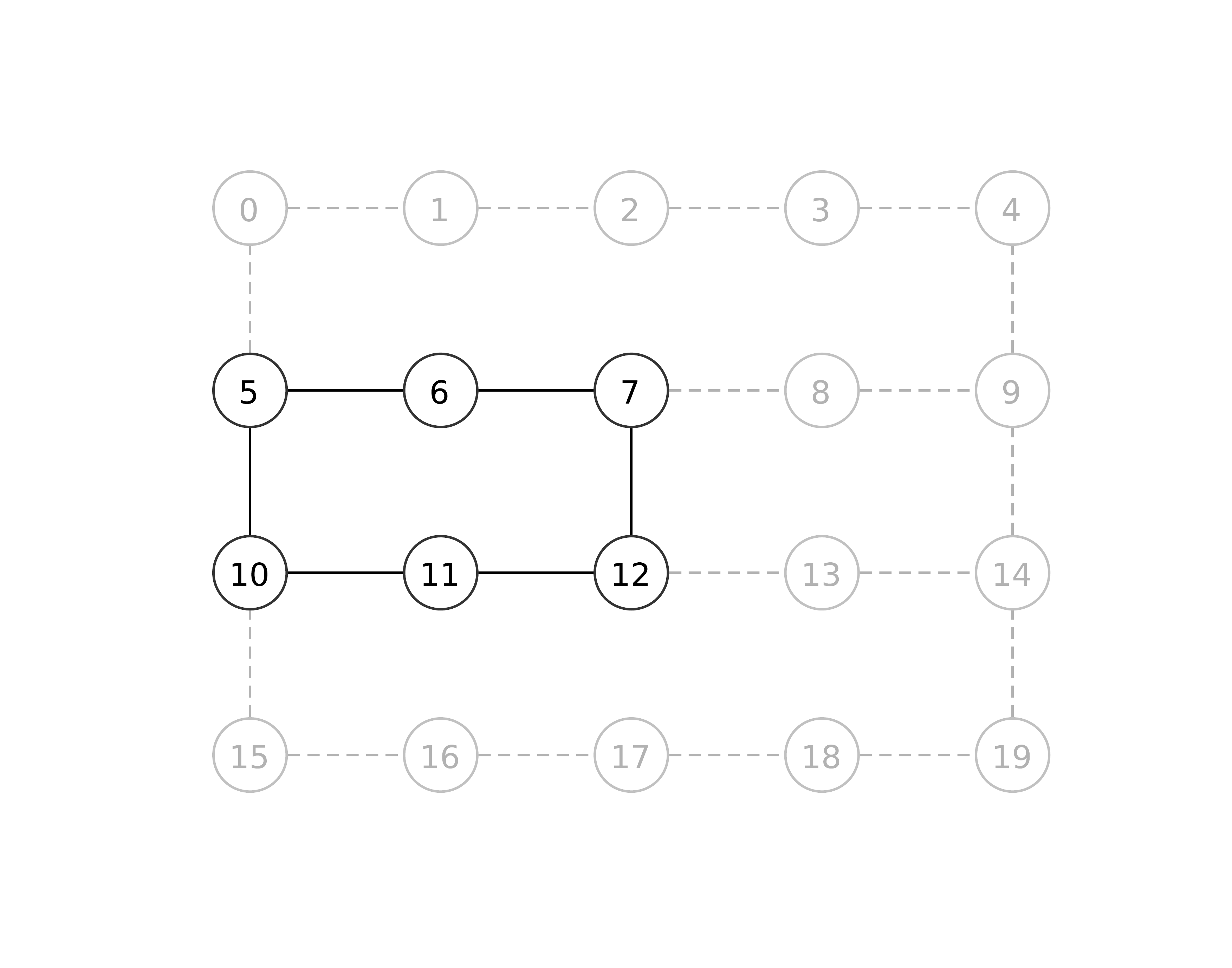}  
\caption{Connectivity of \emph{IBM Q Poughkeepsie}. We use qubits 5, 6, 7, 10, 11, and 12, and entangle nearest neighbors, i.e., we achieve a cyclic entanglement for every layer in the considered variational form.}
\label{fig:poughkeepsie_topology}
\end{figure}

The variational form is constructed as described in Sec.~\ref{sec:vqe} with nearest neighbor entanglement.
The topology of \emph{IBM Q Poughkeepsie}, the selected qubits and the entanglement are illustrated in Fig.~\ref{fig:poughkeepsie_topology}.

In the remainder of this section, we report results for depth $p=0$ and $p=2$; for the overall setup, as well as results for $p=1$, see Section \ref{sec:quantum_results}.

\begin{figure}[ht]
\includegraphics[width=0.5\textwidth]{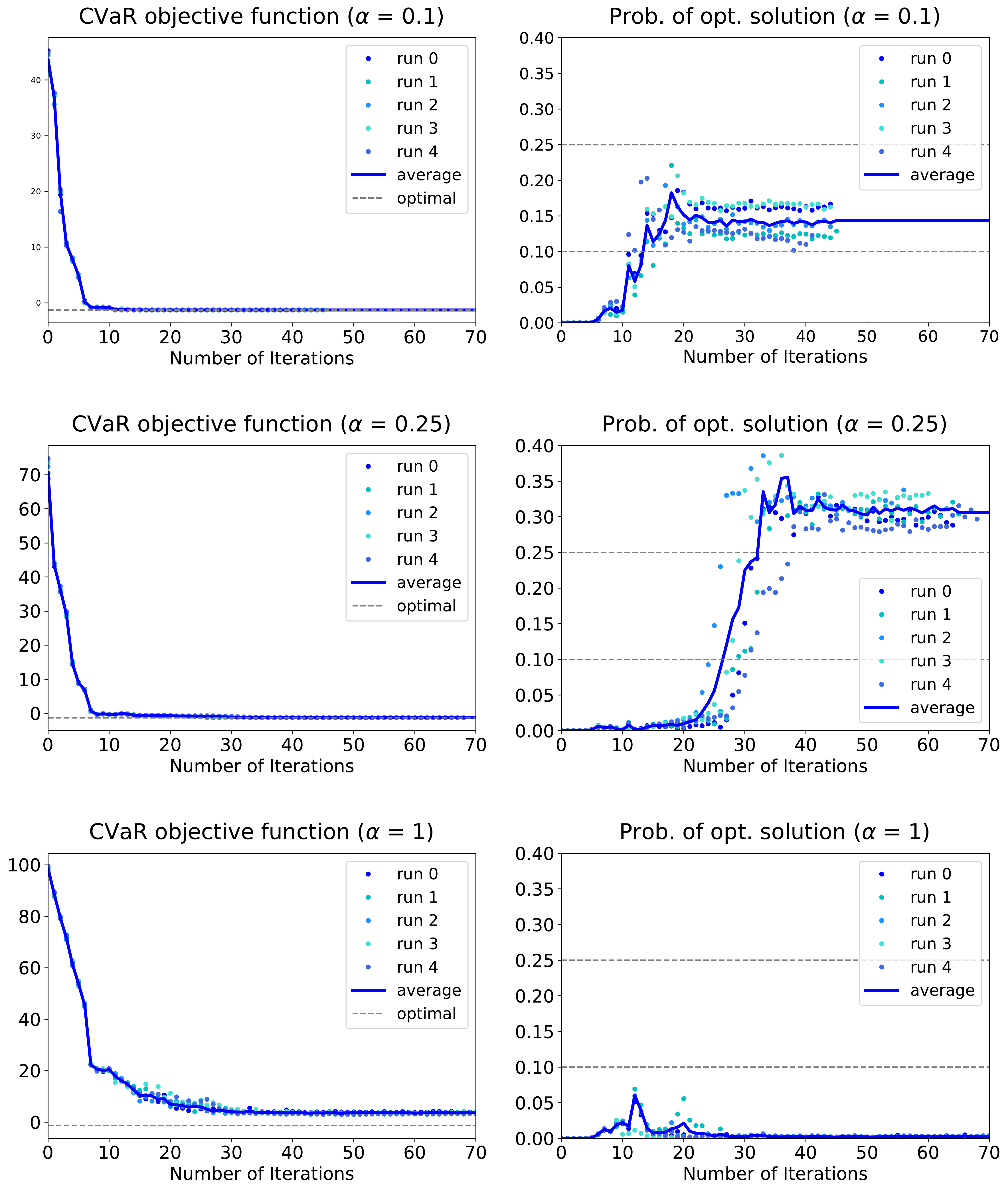}
\caption{Quantum hardware results for portfolio optimization problem
  with six assets/qubits. CVaR-VQE results are shown for depth $p=0$ and
  $\alpha=10\%, 25\%, 100\%$ (from top to bottom) and five runs with
  8,192 samples for each $\alpha$. Plots on the left: resulting objective
  values per iteration; plots on the right: resulting probability of
  sampling an optimal solution. Since COBYLA converges after a
  different number of iterations in each run, we assume that the contribution of each run to the average value after termination of that run is its last reported value. The $\alpha$-levels $10\%$ and $25\%$ are indicated by the
  gray dashed lines in all probability plots.}
\label{fig:real_hardware_results_depth_0}
\end{figure}

For $p=0$,  Fig.~\ref{fig:real_hardware_results_depth_0} shows that results for $\alpha = 10\%$ and $\alpha=25\%$ are similar to those for $p=1$. 
However, for $\alpha=100\%$ the probability of sampling a ground state first increases to $5\%$ on average, then it drops close to zero, even though the objective function improves.
This is an example where improving the objective value does not necessarily imply getting a better overall solution (i.e., binary string), and highlights our motivation of using CVaR as the objective in contrast to the expected value.

\begin{figure}[ht]
\includegraphics[width=0.5\textwidth]{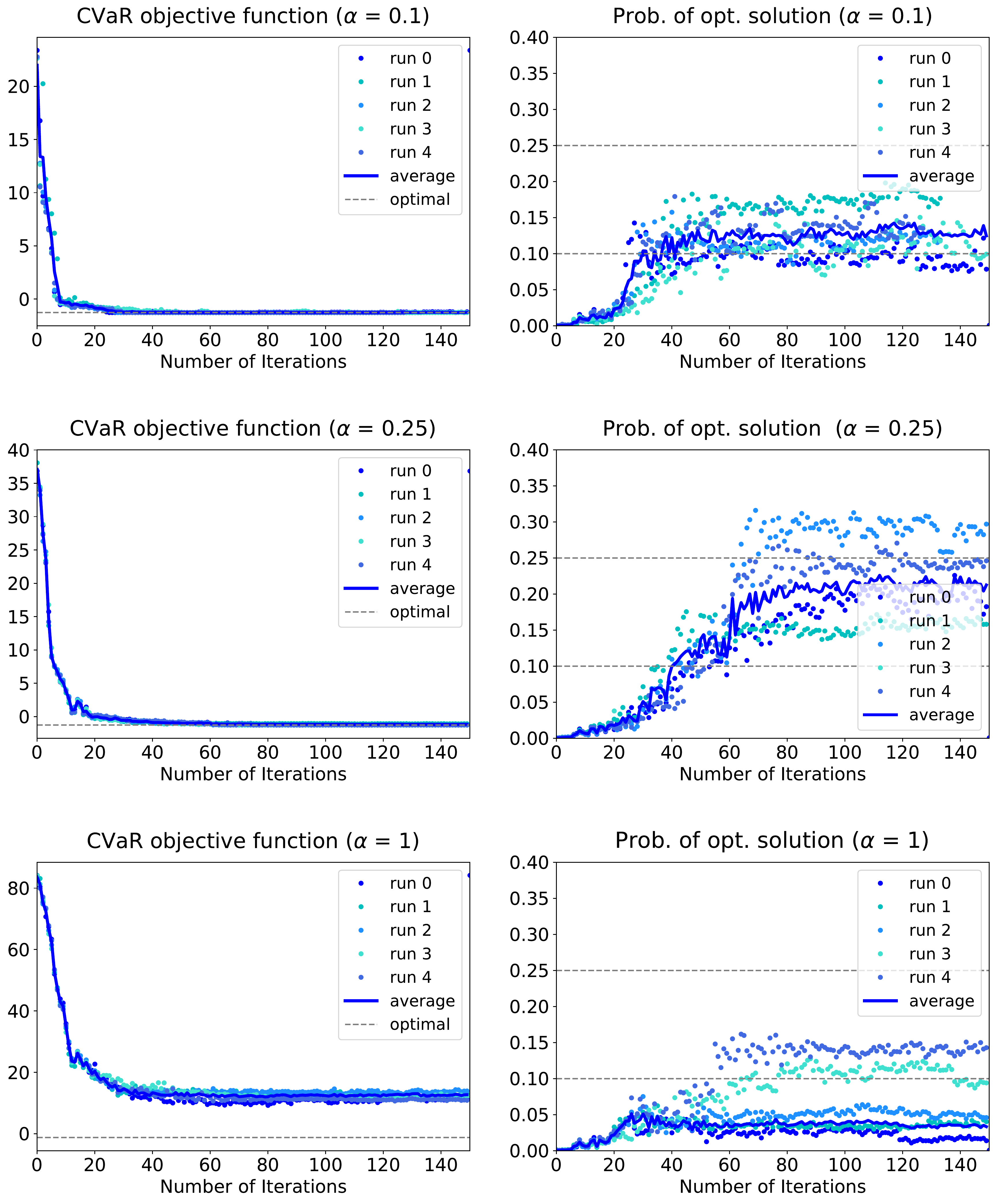}
\caption{Quantum hardware results for portfolio optimization problem
  with six assets/qubits. CVaR-VQE results are shown for depth $p=2$ and
  $\alpha=10\%, 25\%, 100\%$ (from top to bottom) and five runs with
  8,192 samples for each $\alpha$. Plots on the left: resulting objective
  values per iteration; plots on the right: resulting probability of
  sampling an optimal solution.  Since COBYLA converges after a
  different number of iterations in each run, we assume that the contribution of each run to the average value after termination of that run is its last reported value. The $\alpha$-levels $10\%$ and $25\%$ are indicated by the
  gray dashed lines in all probability plots.}
\label{fig:real_hardware_results_depth_2}
\end{figure}

For $p=2$, Fig.~\ref{fig:real_hardware_results_depth_2} again shows that results for $\alpha = 10\%$ are similar to those for $p=0, 1$.
Although the probability of sampling a ground state is not always exceeding $\alpha$ as before, it reaches that level on average.
However, for $\alpha = 25\%$ the probability of sampling a ground state does not reach $\alpha$ anymore, but plateaus slightly below.
For the expected value, i.e., $\alpha=100\%$, we again see a probability of the ground state which is close to zero and that decreases after an initial small increase.

\section{\label{sec:additional_plots} Additional plots}

\begin{figure*}[ht!]
\centering
\includegraphics[width=0.8\textwidth]{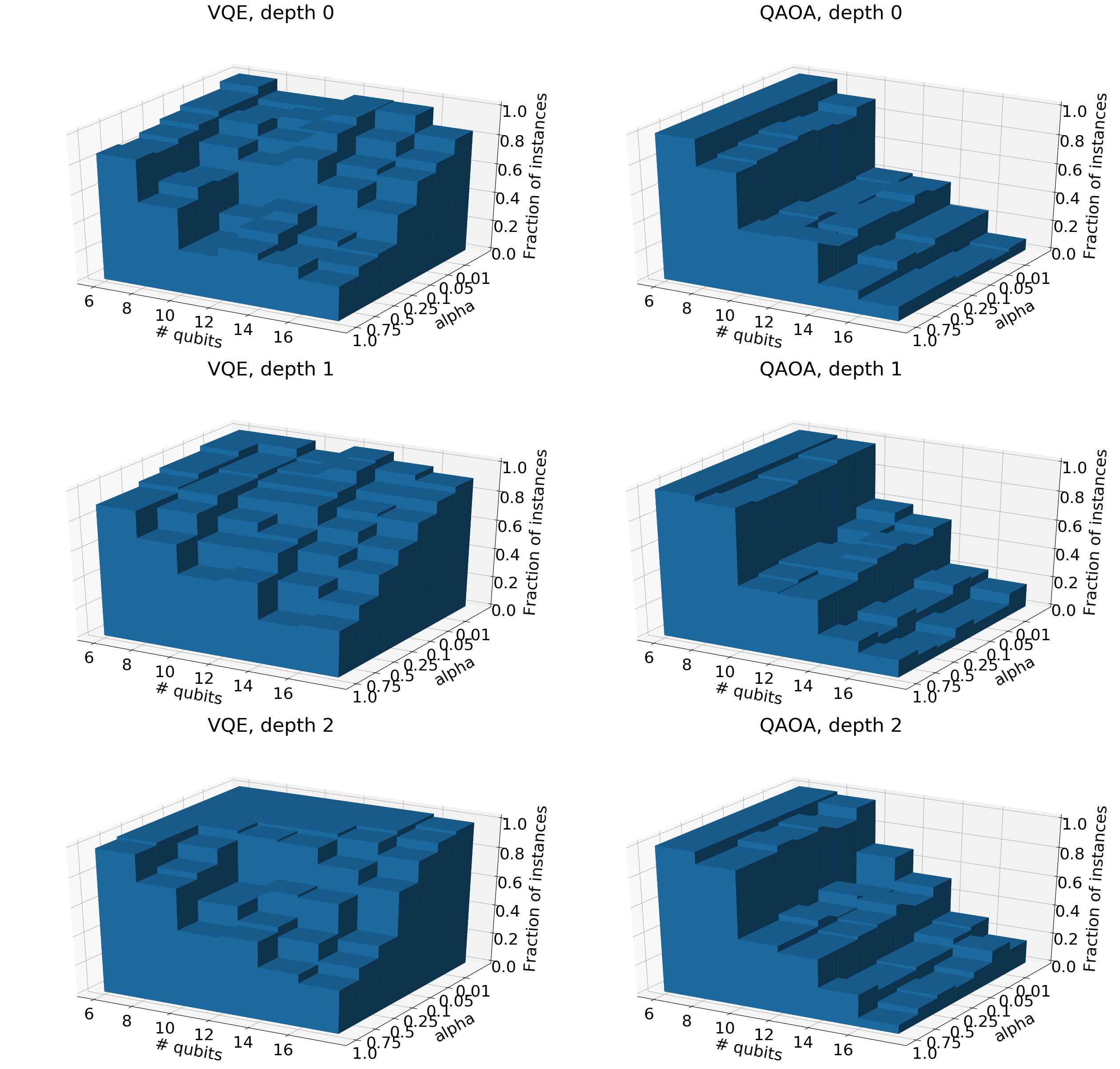}

\caption{
Summary of the results with VQE (left) / QAOA (right) using quantum states resulting from classical simulation, after 50 normalized iterations of the classical optimization algorithm. On the vertical $z$-axis we plot the fraction of instances that attain $1\%$ probability of sampling the optimal solution, on the $xy$-plane we indicate the number of qubits and the value of $\alpha$.}
\label{fig:vqe_qaoa_3d_prob_001}

\end{figure*}

Fig.~\ref{fig:vqe_qaoa_3d_prob_001} shows additional results for VQE and QAOA for different values of $\alpha$, different numbers of qubits, and an overlap with the optimal solution of $1\%$.

\section{\label{sec:proof_qaoa} Proof of Proposition \ref{prop:qaoa}}
We show this by induction on $p$. Let $u_p$ be the $\arg \max$ in the
definition of $\Delta_p$ for $t=p$. With $p=0$, QAOA only applies a
layer of Hadamard gates, therefore it is obvious that $|\alpha_{0,z}|
= \frac{1}{\sqrt{2^n}}$. We now show the induction step.  Recall that
the $p+1$-th layer of QAOA applies two unitaries $U_C(\gamma),
U_B(\beta)$ to the state $\ket{\psi_p}$, in the given order. Here, the
objective function value of a basis state $\ket{z}$ is denoted
$H_{z,z}$ for consistency with the rest of the paper. In classical
QAOA notation, it is typically denoted $C(z) := \sum_{k=1}^{m}
C_k(z)$, with $C_k(z) = 0$ if string $z$ does not satisfy the $k$-th
clause, $C_k(z) = 1$ otherwise, and the goal is to maximize $C(z)$. We can simply think of $H$ as encoding $C$, i.e., $H_{z,z} = C(z)$. Let
$\ket{\phi}$ be the state obtained after applying $U_C(\gamma)$. We
have:
\begin{align*}
  \ket{\phi} &= U_C(\gamma) \ket{\phi_p} =  \sum_{z} \alpha_{p,z} e^{-i \gamma H_{z,z}} \ket{z}.
\end{align*}
Finally we apply $U_B(\beta) = (e^{-i \beta X})^{\otimes n}$ to obtain the state $\ket{\psi} = \sum_{z} \alpha_{p+1,z} \ket{z}$. Let
$e^{-i \beta X} = \begin{pmatrix} a_{00} & a_{01} \\ a_{10} &
  a_{11} \end{pmatrix}$. By definition, for every basis state
$\ket{j}$, we have:
\begin{align}
  \alpha_{p+1,j} =& \sum_{z} \prod_{h=1}^n a_{j_h z_h} \alpha_{p,z} e^{-i \gamma H_{z,z}}  \label{eq:alphaj} \\
  =& \sum_{z} \prod_{h=1}^n  a_{j_h z_h} \alpha_{p,z} e^{-i \gamma \ell} + \notag \\
  & \sum_{z} \prod_{h=1}^n a_{j_h z_h} \alpha_{p,z} e^{-i \gamma \ell}\Big( e^{-i \gamma(H_{z,z} - \ell)} - 1\Big) \notag \\
  =& \sum_{z} \prod_{h=1}^n  a_{j_h z_h} u_p e^{-i \gamma \ell} + \notag\\
  & \sum_{z} \prod_{h=1}^n  a_{j_h z_h} (\alpha_{p,z} - u_p) e^{-i \gamma \ell} + \notag\\
  & \sum_{z} \prod_{h=1}^n a_{j_h z_h} \alpha_{p,z} e^{-i \gamma \ell}\Big( e^{-i \gamma(H_{z,z} - \ell)} - 1\Big)  \notag\\
  =& \big((a_{00} + a_{01})^{\sum_{h} j_h}(a_{10} + a_{11})^{n - \sum_{h} j_h} \notag \\
  & u_p e^{-i \gamma \ell}\big) + \notag\\
  & \sum_{z} \prod_{h=1}^n  a_{j_h z_h} (\alpha_{p,z} - u_p) e^{-i \gamma \ell} + \notag\\
  &\sum_{z} \prod_{h=1}^n a_{j_h z_h} \alpha_{p,z} e^{-i \gamma \ell} \Big( e^{-i \gamma(H_{z,z} - \ell)} - 1\Big). \notag
\end{align}
So
\begin{align*}
  |\alpha_{p+1,j}| \le\\ \left| (a_{00} + a_{01})^{\sum_{h} j_h}(a_{10} + a_{11})^{n - \sum_{h} j_h} u_p e^{-i \gamma \ell}\right| + \\
  \left|\sum_{z} \prod_{h=1}^n  a_{j_h z_h} (\alpha_{p,z} - u_p) e^{-i \gamma \ell}\right| + \\
  \left|\sum_{z} \prod_{h=1}^n a_{j_h z_h}  \alpha_{p,z} e^{-i \gamma \ell} \Big( e^{-i \gamma(H_{z,z} - \ell)} - 1\Big)\right|.
\end{align*}
Recalling that $a_{00} = a_{11} = \cos \beta, a_{01} = a_{10} = -i
\sin \beta$, and $|u_p| \le \max |\alpha_{p,z}|$, it follows that
\begin{align*}
  \left| (a_{00} + a_{01})^{\sum_{h} j_h}(a_{10} + a_{11})^{n -
    \sum_{h} j_h} u_p e^{-i \gamma \ell}\right| \le \\ \max_{z} |\alpha_{p,z}|.
\end{align*}
We also have:
\begin{align*}
  \left|\sum_{z} \prod_{h=1}^n  a_{j_h z_h} (\alpha_{p,z} - u_p) e^{-i \gamma \ell}\right| = \\
  \left|\sum_{z : \alpha_{p,z} \neq u_p} \prod_{h=1}^n  a_{j_h z_h} (\alpha_{p,z} - u_p) e^{-i \gamma \ell}\right| \le \\
  2^n(1-\Delta_p) \left| \prod_{h=1}^n  a_{j_h z_h}\right| \left|(\alpha_{p,z} - u_p)\right| \left|e^{-i \gamma \ell}\right| \le \\
  2^n(1-\Delta_p) (\left|\alpha_{p,z}\right| + \left| u_p \right|) \le\\ 2^n(1-\Delta_p)(2 \max_{z} |\alpha_{p,z}|). 
\end{align*}
We now need to find an upper bound to the term $\left|\sum_{z}
\prod_{h=1}^n a_{j_h z_h} \alpha_{p,z} e^{-i \gamma \ell} (e^{-i \gamma(H_{z,z} - \ell)} - 1)\right|$. We obtain:
\begin{align*}
  \left|\sum_{z} \prod_{h=1}^n a_{j_h z_h} \alpha_{p,z} e^{-i \gamma \ell} (e^{-i \gamma(H_{z,z} - \ell)} - 1)\right|
  \le \\ \sum_{z} \prod_{h=1}^n \left| a_{j_h z_h} \alpha_{p,z} e^{-i \gamma \ell}\right| \left|(e^{-i \gamma (H_{z,z} - \ell)} - 1)\right| \le \\
  \sum_{z} |\alpha_{p,z}| \left|(e^{-i \gamma (H_{z,z} - \ell)} - 1)\right| \le \\
  \max_{z} |\alpha_{p,z}| \sum_{z : H_{z,z} = \ell}  \left|(e^{-i \gamma (\ell - \ell)} - 1)\right| + \\ \max_{z} |\alpha_{p,z}| \sum_{z : H_{z,z} \neq \ell} \left| e^{-i \gamma (H_{z,z} - \ell)} - 1 \right|\le \\
  2\max_{z} |\alpha_{p,z}| 2^n (1-\delta)
\end{align*}
Thus, we have
\begin{align*}
  |\alpha_{p+1,j}| &\le (2^{n+1} (2 - \Delta_p -\delta ) + 1)\max_{z} |\alpha_{p,z}| \\
  &\le (2^{n+1} (2 - \Delta_p -\delta ) + 1)^{p+1} \frac{1}{\sqrt{2^n}},
\end{align*}
where we used the fact that $\Delta_p$ is decreasing in $p$ by
definition.  It is also useful to determine a lower bound on
$\Delta_p$. It is clear that $\Delta_0 = 1$ because for $p = 0$ all
amplitudes are equal. To find a lower bound on $\Delta_{p+1}$ based on
$\Delta_p$, we look at the last line of \eqref{eq:alphaj}, which
decomposes $\alpha_{p+1,j}$ into three summations. The first summation
has $n$ possible different values. The second summation has at most
$\frac{n^2}{4}$ possible coefficient values for each $z$: this is
because $\prod_{h=1}^n a_{j_h z_h}$ can be computed by looking at
which $z_h$ are 1 and counting how many corresponding $j_h$ are 1,
then doing the same for zeros. The largest number of combinations is
obtained when $z$ has $n/2$ bits equal to 1, yielding $(n/2)^2$
combinations. Since there are $(1-\Delta_p)2^n$ nonzero terms in the
summation, in total we obtain at most $(n^2/4)^{(1-\Delta_p)2^n}$
different values. The third summation is similar: $\frac{n^2}{4}$
possible coefficient values for each $z$, and $(1-\delta) 2^n$ nonzero
terms, for a total of at most $(n^2/4)^{(1-\delta) 2^n}$ different
values. In total, there are at most
$n(n^2/4)^{(2-\delta-\Delta_p)2^n}$ different values of
$\alpha_j$. Hence, $\Delta_{p+1} \ge
1/(n(n^2/4)^{(2-\delta-\Delta_p)2^n})$. With algebraic manipulations,
we obtain a (possibly very loose) bound $\Delta_{p} \ge
(\frac{1}{n^3})^{2^n(1 - \delta + \frac{p-1}{p})}$ whenever $p \ge 1$.

\bibliographystyle{plainnat}
\bibliography{bibliography}

\end{document}